\documentclass[reprint,superscriptaddress,aps,floatfix,prb,amsmath,amssymb]{revtex4-2}
\usepackage[utf8]{inputenc}
\usepackage{bm}
\usepackage{graphicx}
\usepackage{hyperref}
\usepackage{color}
\usepackage{nicefrac}
\usepackage{booktabs}
\usepackage{natbib}

\begin{document}

\title{Trilinear coupling driven ferroelectricity in HfO$_2$}

\author{Francesco Delodovici}
\affiliation{Consiglio Nazionale delle Ricerche CNR-SPIN, c/o Universit\`a degli Studi “G. D’Annunzio”, I-66100 Chieti, Italy}

\author{Paolo Barone}
\affiliation{Consiglio Nazionale delle Ricerche CNR-SPIN, Area della Ricerca di Tor Vergata, Via del Fosso del Cavaliere, 100, I-00133 Rome, Italy}

\author{Silvia Picozzi}
\affiliation{Consiglio Nazionale delle Ricerche CNR-SPIN, c/o Universit\`a degli Studi “G. D’Annunzio”, I-66100 Chieti, Italy}

\begin{abstract}
Ferroelectricity in hafnia is often regarded as a breakthrough  discovery in  ferroelectrics, potentially able to revolutionize the whole field. Despite increasing interests, a deep and comprehensive understanding of the many factors driving the ferroelectric stabilization is still lacking. We here address the phase transition in terms of a Landau-theory-based approach, by  analyzing  symmetry-allowed distortions connecting the high--symmetry paraelectric tetragonal phase to the low--symmetry polar orthorhombic phase. By means of first-principles simulations, we find that the  $\Gamma_{3-}$ polar mode is only weakly unstable, whereas the other two symmetry--allowed distortions, Y$_{2+}$ and Y$_{4-}$ (showing a non-polar and antipolar behaviour, respectively) are hard modes.
While none of the modes, taken alone or  combined with one other mode, is able to drive the transition,  the key factor in stabilizing the ferroelectric phase is identified as the strong trilinear coupling among the three modes. Furthermore, the experimentally acknowledged importance of substrate-induced effects in the growth of  HfO$_2$ ferroelectric  thin films, along with the 
lack of a clear order parameter in the transition, suggested the extension of our analysis to strain effects. Our findings suggest a complex behaviour of the  Y$_{2+}$ mode, which can become unstable under certain conditions ({\em i.e.} a tensile strain applied along the {\bf a} direction) and an overall weakly unstable behaviour for the $\Gamma_{3-}$ polar mode for all the strain conditions. In any case, a robust result emerges from our analysis:
 independently of the different applied strain 
(be it compressive or tensile, applied along  {\bf a,b } or {\bf c} orthorhombic axes), the need of a simultaneous excitation of the three coupled modes remain unaltered. Finally, when applied to mimic experimental growth conditions under strain, our analysis  show a further stabilization of the ferroelectric phase with respect to the unstrained case, in agreeement with experimental findings.
\end{abstract}

\maketitle

\section{Introduction}
\large
Hafnium dioxide has been widely studied in its non-polar phases as a high-dielectric-constant non-toxic oxide with crucial exploitation in microelectronics \cite{Lin_2002}. 
In fact, hafnia has been heavily employed as a gate dielectric in field-effect transistors, due to its compatibility with Si \cite{Wilk_2000,Zhu_2012,Robertson_2005} 
and as a favorable gate dielectric in  metal-oxide-semiconductor field-effect transistors. 
In 2011 the discovery of its ferroelectric phase in thin films \cite{Schroder_2011,Schroder_2011_Proceedings}
shed a light on a new intriguing phenomenon and opened additional and promising avenues for HfO$_2$-based applications. 
In particular, due to their switchable persistent out-of-plane polarization, hafnia films could be used in high-speed non-volatile memories and logic devices, such as ferroelectric field-effect transistors \cite{Mikolajick_2018} 
and ferroelectric tunnel junctions \cite{Noheda_2019_tunnJunc}. 
Even though different polar phases have been detected 
\cite{Barabash_2017,Noheda_2018,Qi_2020}, 
the polar orthorhombic phase $Pca2_1$ is mostly recognized to be responsible for ferroelectricity \cite{Schroder_2011}. 
In bulk configuration, this phase is metastable at standard thermodynamical conditions, where the lowest energy phase is the non-polar monoclinic $P2_1c$ \cite{Huan_2014,Schroder_2011}; 
as such, the pristine ferroelectric phase cannot be stabilized through simple thermodynamical transformations.
In fact, Y-doped $Pca2_1$ phase has been recently obtained in bulk form through an ultra-rapid cooling following high-T annealing \cite{Xu_2021}. 
On the other hand, when the thin-film limit is approached, the orthorhombic polar phase appears to be energetically favoured, becoming naturally competitive with the monoclinic non-polar phase \cite{Tsymbal_2020}. 
This condition is robust against the decrease of film thickness down to subnanometer-scale \cite{Yurchuk_2013,Lee_2020_SCIENCE}, 
at variance with the usual behaviour observed in perovskites-based thin films \cite{junquera_2003,Batra_1973,Ma_2002,Dawber_2005}. 
Indeed, as the film thickness decreases, the depolarization field produced by the incomplete screening of surface charges is expected to increase, challenging the stabilization of the out-of-plane switchable polarization.
Nonetheless, stable ferroelectric HfO$_2$ nanometer-thick films with finite out-of-plane polarization are regularly grown on different substrates \cite{Estandia_2019,Estandia_2020,Noheda_2018,Noheda_2019}, 
a characteristic which further increases their potentiality in electronics.
The mechanism causing the formation of the polar phase in films is probably dependent on the growth method, but evidences exist \cite{Schroder_2011,Muller_2012,Grimley_2018,Estandia_2019} 
revealing the tetragonal phase as possible precursor for the activation of ferroelectricity, as the transition temperature from the monoclinic phase is suppressed approaching low dimensions \cite{Muller_2012} . 

Despite the large amount of analysis, to the best of our  knowledge, a clear microscopic description of the mechanism underlying the stabilization of ferroelectricity in hafnia is still missing. 
At a phenomenological level this mechanism is usually recognized as a superposition of different effects.
Possible invoked causes are  doping  \cite{Batra_2017,Schroder_2018}, 
oxygen vacancies \cite{hoffmann_2015}, 
grains size and the films thickness  \cite{Polakowski_2015} . 
Recently the scale-free nature of ferroelectricity in hafnia and its high coercive field have been connected to intrinsic flat phonon-bands \cite{Lee_2020_SCIENCE}.

Another important factor is represented by mechanical effects induced by the substrate \cite{Batra_2017,Estandia_2019,Estandia_2020,Liu_2019}. 
In closer detail, the external strain plays a major role in the stabilization of the orthorhombic phase in Hafnia thin films, as induced by the lattice mismatch with the  substrate.
As confirmed by experiments \cite{Estandia_2019} 
and simulations \cite{Tsymbal_2020}, 
strain changes the relative amount of competing phases present in the films, favouring either the polar orthorhombic or the non-polar monoclinic phase depending on the lattice constant of the underlying substrate.

In this paper we present a first--principles--based study of the phase transition from the bulk non-polar tetragonal $P4_2nmc$ to the bulk polar orthorhombic phase $Pca2_1$ and interpret our results according to a symmetry distortion mode analysis. 
We further include external strains applied to the HfO$_2$ primitive cell, in order to study the dependence of the distortion modes on external mechanical actions, in the aim of understanding how certain strain states may affect the phase transition.
Finally, we address an experimentally relevant configuration as a benchmark of our analysis.

\section{Symmetry mode analysis}
The transition from the high-symmetry non-polar tetragonal phase ($P4_2nmc$) to the polar orthorhombic phase ($Pca2_1$) couples atomic displacements with a non-trivial deformation of the cell. 
The analysis of the symmetry-allowed distortions \cite{Stokes_2006,Stokes_2016,Orobengoa_2009,Perez-Mato_2010} 
reveals five patterns connecting the two phases: three modes at Brillouin zone center $\Gamma_{1+}$, $\Gamma_{4+}$ $\Gamma_{5-}$, and two zone-boundary modes M$_{1}$ and M$_{3}$.
The $\Gamma_{1+}$ involves atomic displacements coupled to a volume expansion, preserving the tetragonal symmetry.
The $ \Gamma_{4+}$ mode is a pure-strain, volume-conserving mode responsible for lowering the tetragonal symmetry to the orthorhombic crystal class: it corresponds to a small distortion of the angle $\gamma$ from its ideal tetragonal value to the low-symmetry relaxed value of 90.25$^{\circ}$. 
\begin{figure}[b]
\centerline{
\includegraphics[angle=0,width=6cm]{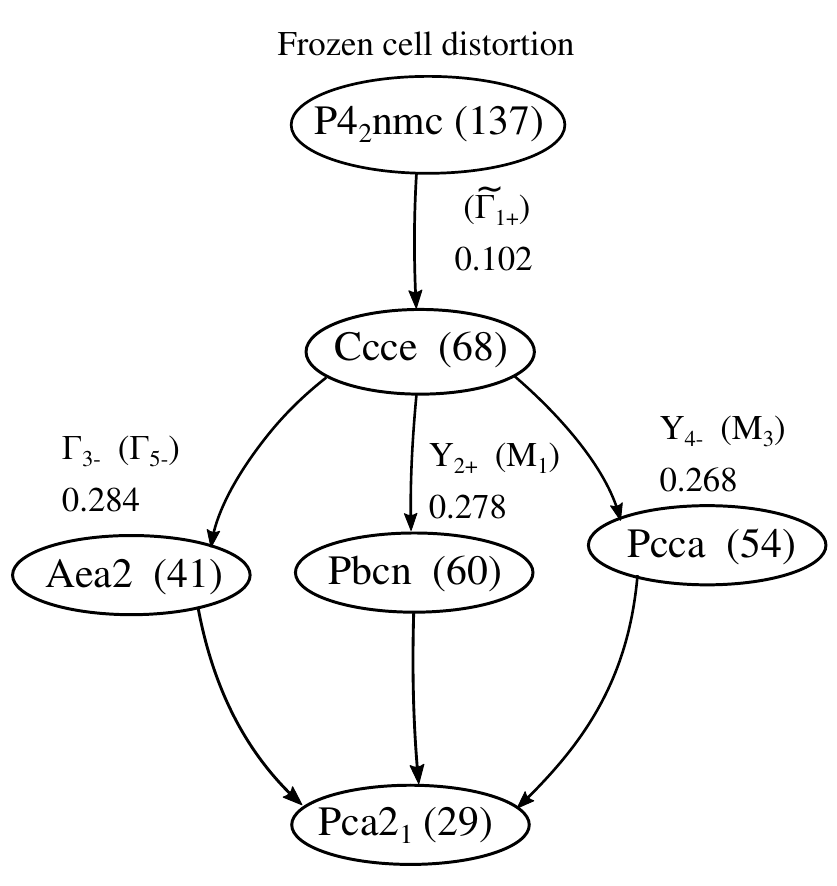}
}
\caption{ \label{pca21_symmetry_tree} 
   The super-group tree of the orthorhombic $Pca2_1$ structure. The parent structure $P4_2nmc$ distorted into the $Pca2_1$ primitive cell is connected to the $Ccce$ structure. The space-group number is reported in parenthesis. For each distortion, the amplitude of the maximum atomic displacement is reported in $\rm \AA$ below the corresponding irreps. The displacements are intended from the closest high-symmetry phase atomic coordinates.
}
\end{figure}
The full orthorhombic cell is obtained via the doubling of this resulting distorted tetragonal cell. 
In order to simplify the description, still retaining the main ingredients, we here focus only on the  contributions to the phase transition given by atomic-displacements, {\em i.e.} throughout our analysis we fixed the volume and shape of the cell to those of the orthorhombic primitive cell .
In closer detail, this choice corresponds to freeze the $\Gamma_{4+}$ and the normal-strain component of $\Gamma_{1+}$.
\begin{figure}[]
\centerline{
\includegraphics[angle=-90,width=0.52\textwidth]{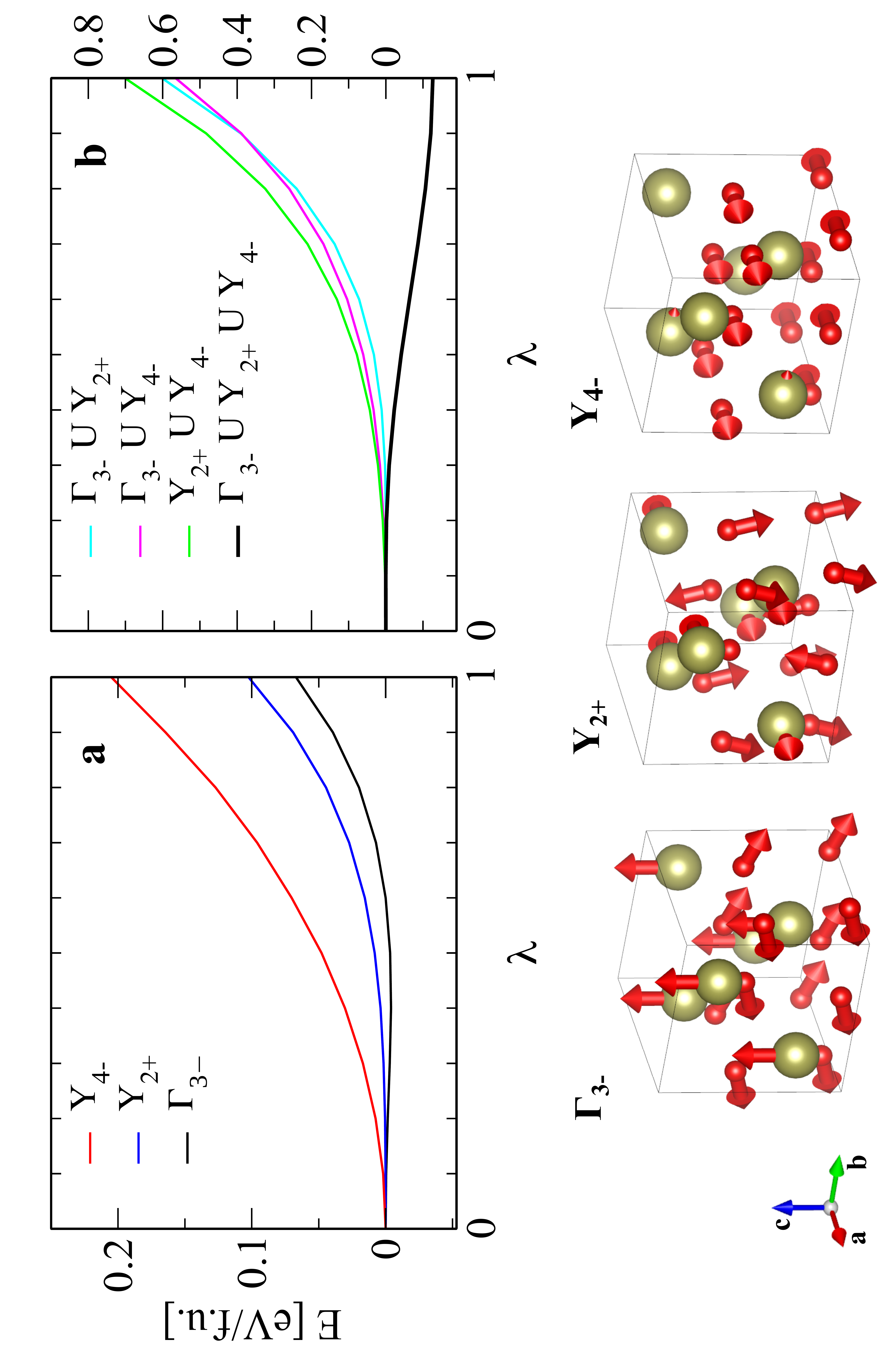}
}
\caption{ \label{panel}
    Top: {\it ab-initio} energies (in eV/f.u.) of the single symmetry-distortion modes (panel a), along with their couplings (panel b), as a function of the generalized distortion coordinate $\lambda$. Each path consists of 10 intermediate structures. The energy trend relative to the simultaneous excitation of the three modes is reported with a bold black line. 
    Bottom: atomic displacements of each single mode. Hafnium is colored in gold, and oxygen in red. The arrows point in the direction of the displacement with a length proportional to its magnitude. For every mode the arrows relative to hafnium are magnified by a factor of 5.
}
\end{figure}

Since the atomic part of $\Gamma_{1+}$ ($\tilde{\Gamma}_{1+}$ in Fig. \ref{pca21_symmetry_tree}) does not lower the symmetry of the atomic internal coordinates, one is allowed to freeze it at its maximum amplitude, taking the resulting $Ccce$ configuration (space group 68) as the high-symmetry structure for our analysis.
We notice that {\em ab initio} calculations based on Density Functional Theory (DFT) reveal that the total energy of such idealized structure  is  48 meV/f.u. higher than the energy of the tetragonal phase in its optimized volume; still, this difference is smaller than the energy separation between the tetragonal and the cubic phase (61 meV/f.u. above the tetragonal). 
Nonetheless, the choice of the high-symmetry parent structure in the orthorhombic class unveils some interesting features of the phase transition.
The distortion-symmetry modes connecting the $Ccce$ to the $Pca2_1$ structure, reported in Fig.\ref{pca21_symmetry_tree} labelled according to the $Ccce$ irreps, drives the system in three different intermediate states, namely: the $Aea2_1$ (41), $Pbcn$ (60) and $Pcca$ (54), connected to the parent structure (68) respectively through the modes $\Gamma_{3-}$ , Y$_{2+}$ and Y$_{4-}$.
We note that these three distortions correspond respectively to the $\Gamma_{5-}$ , M$_1$ and M$_3$ using the labelling of the tetragonal irreps. This can be verified by comparing the intermediate states produced by the activation of the single modes.
$\Gamma_{3-}$ is a polar mode, while Y$_{2+}$ and Y$_{4-}$ are non-polar and anti-polar modes, respectively, at the Brillouin zone boundary.
%
The atomic displacements characteristic of each mode is reported in the lower half of Fig.\ref{panel}. The oxygens rotate around the \textbf{a} primitive vector in the Y$_{2+}$ mode whereas they undergo an anti-phase shift parallel to the same axis \textbf{a} in the Y$_{4-}$. 
%
The overall effect of $\Gamma_{3-}$ is to move apart the oxygens and the hafnium atoms along the \textbf{c} primitive vector, which causes the non-vanishing polarization.

Fig. \ref{panel}.a reports the trend of the {\it ab-initio} total energy as a function of a $\lambda$ generalized coordinate representing the single mode distortion. We identify Y$_{2+}$ and Y$_{4-}$ as hard modes and $\Gamma_{3-}$ as a soft mode connected to a weak instability (4 meV/f.u.).
Despite this small instability, the polar mode does not drive by itself the transition to the ferroelectric phase. Instead it leads towards the metastable $Aea2_1$ phase, which is 193 meV/f.u. higher in energy with respect to the $Pca2_1$.
We thus analyze the {\it ab-initio} total energies for the coupling between modes (reported in Fig. \ref{panel} b). 
The energy trends reveal the stability of the $Ccce$ structure against all pair-wise combinations of modes; 
in other words,  the hardness of both 
non-polar modes suppresses the $\Gamma_{3-}$ weak instability when combined to the polar mode.
%
Besides, the two-mode couplings strengthen the stability of $Ccce$, as can be seen by the difference in the energy scale between panel a) and b) in Fig. \ref{panel}.
The continuous black line in Fig. \ref{panel}.b corresponds to the trend of the three modes coupling. The final structure at $\lambda = 1$ corresponds to an energy gain of 0.126 eV/f.u. and shows the $Pca2_1$ symmetry group \cite{Stokes_2005,Findsym}. 
This confirms the necessity of the simultaneous excitation of the three modes to stabilize the ferroelectric phase, since the polar $\Gamma_{3-}$ mode is unable  - neither by itself  nor combined with a single hard mode - to provide the required energy gain. 
The possible role played by trilinear coupling in inducing ferroelectric transitions has been already highlighted in so-called hybrid improper ferroelectrics \cite{Bousquet2008, BenedekFennie2011,Stroppa2013,Iniguez2014}. 
In these systems, the ferroelectric transition is mainly driven by the hybrid mode resulting from the coupling of two non-polar distortions, at odds with the situation realized in HfO$_2$, where the coupled $Y_{2+}\,\cup\,Y_{4-}$ is in fact the hardest one (see Fig. \ref{panel}.b).
In order to gain more insights on the modes coupling driving the phase transition, we follow a Landau-theory approach and analyze the free energy landscape surrounding the $Ccce$ structure as a function of its symmetry-allowed distortion modes.
The free energy expanded up to the sixth order in the order parameter describing each distortion is:
\begin{widetext}
\begin{eqnarray}
E(Q_{\Gamma_{3-}},Q_{Y_{2+}},Q_{Y_{4-}}) &=& E_0 + \beta_{200}Q^2_{\Gamma_{3-}}+\beta_{020}Q^2_{Y_{2+}}+\beta_{002}Q^2_{Y_{4-}} + \gamma_{111}Q_{\Gamma_{3-}}Q_{Y_{2+}}Q_{Y_{4-}} + \delta_{400}Q^4_{\Gamma_{3-}} +\nonumber\\
&+&\delta_{040}Q^4_{Y_{2+}}+\delta_{004}Q^4_{Y_{4-}} +\delta_{220}Q^2_{\Gamma_{3-}}Q^2_{Y_{2+}}+\delta_{202}Q^2_{\Gamma_{3-}}Q^2_{Y_{4-}}+\delta_{022}Q^2_{Y_{2+}}Q^2_{Y_{4-}}+\nonumber\\
        &+&\epsilon_{311}Q^3_{\Gamma_{3-}}Q_{Y_{2+}}Q_{Y_{4-}} + \epsilon_{131}Q_{\Gamma_{3-}}Q^3_{Y_{2+}}Q_{Y_{4-}}+\epsilon_{113}Q_{\Gamma_{3-}}Q_{Y_{2+}}Q^3_{Y_{4-}} + \nonumber\\
        &+&\eta_{600}Q^6_{\Gamma_{3-}}+\eta_{060}Q^6_{Y_{2+}}+\eta_{006}Q^6_{Y_{4-}}+\eta_{420}Q^4_{\Gamma_{3-}}Q^2_{Y_{2+}} + \eta_{240}Q^2_{\Gamma_{3-}}Q^4_{Y_{2+}} + \nonumber \\  &+&\eta_{402}Q^4_{\Gamma_{3-}}Q^2_{Y_{4-}} +  \eta_{204}Q^2_{\Gamma_{3-}}Q^4_{Y_{4-}} + \eta_{042}Q^4_{Y_{2+}}Q^2_{Y_{4-}} + \eta_{024}Q^2_{Y_{2+}}Q^4_{Y_{4-}} + \nonumber \\
        &+&\eta_{222}Q^2_{\Gamma_{3-}}Q^2_{Y_{2+}}Q^2_{Y_{4-}} + ...
\label{landau}
\end{eqnarray}
\end{widetext}

The order parameters $\rm Q_{\Gamma_{3-}}$, $\rm Q_{Y_{2+}} $, $\rm Q_{Y_{4-}}$ represent the amplitude (dimensionless) of the three modes. $\beta, \gamma, \delta, \epsilon, \eta$ are the expansion coefficients (in eV/f.u.). 
The condition $\rm Q = (Q_{\Gamma_{3-}}, Q_{Y_{2+}}, Q_{Y_{4-}})=(0,0,0)$ corresponds to the high-symmetry $Ccce$, whereas $\rm Q=(1,1,1)$ corresponds to the orthorhombic $Pca2_1$.
The free energy includes coupling at different orders: trilinear $\gamma$, biquadratic $\delta$, and some more complex term i.e. bilinear-cubic $\epsilon$ and quadratic-quartic $\eta$. 
The odd terms in equation \ref{landau}, the trilinear and the bilinear-cubic terms, prevent the degeneracy of the $Pca2_1$ against the simultaneous inversion of each mode:
thus the phases $\rm Q=(-1,-1,-1)$ and $\rm Q=(1,1,1)$ are not energetically equivalent.
Instead, the $Pca2_1$ is four fold degenerate in total energy against the simultaneous inversion of couples of modes. Indeed the phases $\rm Q=(1,1,1),\, Q=(1,-1,-1),\, Q=(-1,1,-1),\, Q=(-1,-1,1)$ are all energetically equivalent.
Furthermore, the trilinear terms set the polarity of the pseudo-rotations combination against an inversion of the path: $Y_{2+}\,\cup\,Y_{4-}$ must have the same parity of $\Gamma_{3-}$ mode, which means that the coupling of the rotations corresponds to a polar distortion, as it happens in hybrid improper ferroelectrics \cite{Bousquet2008,BenedekFennie2011,Stroppa2013,Iniguez2014},  
with the crucial difference that such hybrid polar mode does not soften in HfO$_2$ .

\begin{figure}[]
\centerline{
\includegraphics[angle=0,width=7cm]{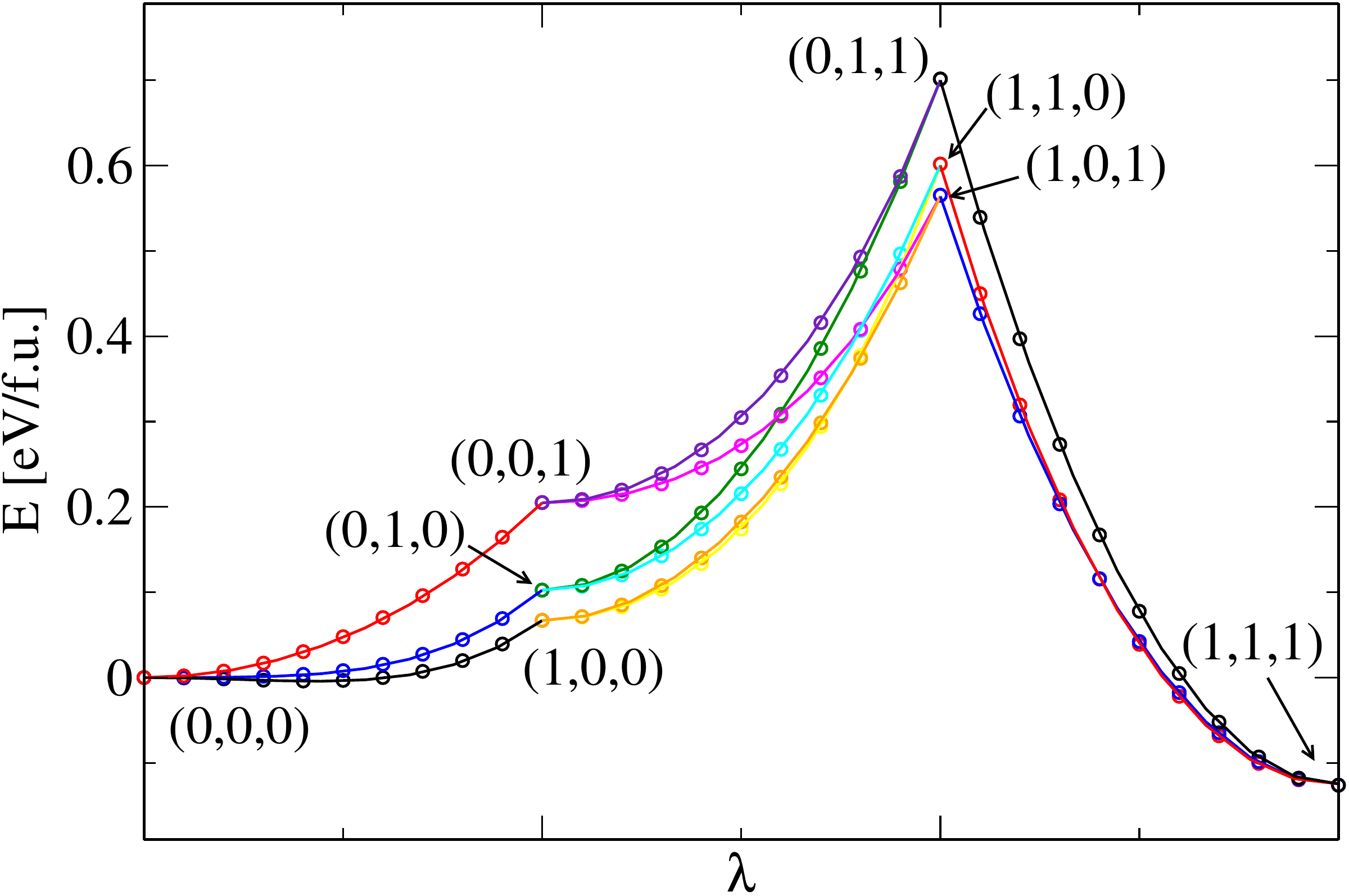}
}
\caption{ \label{fit} 
    Fit of  {\it ab-initio} energies, represented by colored dots, with the free energy expressed in Qq.\ref{landau} for the zero-strain case. The path connects the high symmetry structures from the $Ccce$ to the $Pca2_1$. The figure shows only  part of the energy landscape, {\em i.e.} it reports only some portions of the fit: two-modes and three-modes combination are not shown.
      Each high symmetry structure is labelled with the values of the vector $\rm (Q_{\Gamma_{3-}},Q_{Y_{2+}},Q_{Y_{4-}})$, thus $\rm(0,0,0)$ corresponds to the $Ccce$ and $\rm(1,1,1)$ to the $Pca2_1$. 
      %
    }
\end{figure}

Based on these consideration we focus on the patterns connecting $\rm Q=(0,0,0)$ to $\rm Q=(1,1,1)$ via the high symmetry structures produced by the three single modes and their couplings.
Each portion of the path consists of 10 intermediate structures.
In practice, we first computed the {\it ab-initio} total energies of  different structures along these paths.  
We then extracted a numerical estimate of the Landau coefficients through a fit of the total energies with the polynomial in equation \ref{landau}.
From the ratio of the coefficients we deduced the terms leading the transition and the character of the different contribution to the free energy.
Figure \ref{fit} reports the  {\it ab-initio} energies and the relative fit for the case where no strain is applied. 
The initial portion of the path coincides with the trends in Fig. \ref{panel}.a.
\begin{table}[]
    \centering
    \caption{Fit coefficients obtained by interpolation of the {\it ab-initio} total energies with equation \ref{landau} for the case of zero-applied strain.} 
    \label{coeff_tab}
    \vspace{0.5cm}
    \begin{tabular}{c c c c}
    \toprule
    \toprule
      Coeff.  & [eV/f.u.]  & Coeff.   & [eV/f.u.] \\
    \midrule
    $\beta_{(200)}$ & $-0.046 $ & $\epsilon_{(311)}$ & $-0.147 $ \\
    $\beta_{(020)}$ & $0.001 $ & $\epsilon_{(131)}$ & $-0.372 $ \\
    $\beta_{(002)}$ & $0.180 $  &  $\epsilon_{(113)}$ & $-0.238 $ \\
    $\gamma_{(111)}$ & $-1.080 $ & $\eta_{(600)}$ & $-0.011 $ \\
    $\delta_{(400)}$ & $0.124 $ & $\eta_{(060)}$ & $-0.010 $ \\
    $\delta_{(040)}$ & $0.111 $ & $\eta_{(006)}$ & $-0.005 $ \\
    $\delta_{(004)}$ & $0.029 $ & $\eta_{(420)}$ & $-0.050 $ \\
    $\delta_{(220)}$ & $0.444 $ & $\eta_{(402)}$ & $0.007 $ \\
    $\delta_{(202)}$ & $0.264 $ & $\eta_{(240)}$ & $0.036 $ \\
    $\delta_{(022)}$ & $0.355 $ & $\eta_{(042)}$ & $0.026 $ \\
     & & $\eta_{(024)}$ & $0.012 $ \\
     & & $\eta_{(204)}$ & $0.021 $ \\
     & & $\eta_{(222)}$ & $0.222 $ \\
   \bottomrule
   \bottomrule
    \end{tabular}
\end{table}
The two pseudo-rotations increase the energy of $Ccce$ in a range of 0.1-0.2 eV/f.u. with a monotonous trend: this behaviour is captured by the positiveness of $\beta_{020},\,\beta_{002}$ and $\delta_{040},\,\delta_{004}$.
Interestingly, among all the parameters, the trilinear coupling is  by far the dominant term: $\gamma_{111}\approx-1.08$ eV/f.u, confirming that the three distortions are all equally necessary to complete the phase transition when no strain is present. 
It should be stressed that $\beta_{020}\approx1\,$ meV/f.u. is remarkably smaller than all the other parameters as can be seen in Tab. \ref{coeff_tab}.
This is interesting since a change in the sign of this parameter corresponds to the insurgence of an instability in Y$_{2+}$. This may occur by means of some external perturbation, e.g. mechanical strain, which modifies the relative stability of the atomic configurations along the distortion path. 
Despite the corresponding sixth order $\eta$ are negative, they are significantly smaller than the $\delta$, thus they do not alter the stable character of the rotations.
Nonetheless, such terms are necessary to obtain a reliable fit, especially for the Y$_{2+}$ distortion.
Instead, the coefficient of the  polar mode $\beta_{200},\,\delta_{200}$ have an opposite sign, reflecting the instability of the $\Gamma_{3-}$ mode, evaluated as 3.9 meV/f.u. .
The $\eta_{600}$ is negative and non negligible (albeit small) and indeed it is necessary to obtain a reliable fit.

\section{Strain effects}

As mentioned in the introduction, substrate-induced strain is  widely recognized as a major cause for the stabilization of ferroelectricity in hafnia \cite{Estandia_2019,Estandia_2020,Noheda_2018,Tsymbal_2020}. 
Here we investigate whether 
such an energetically favorable dependence on strain emerges  in the depicted framework, when analyzing the symmetry-allowed distortions.
We recall that the growth of HfO$_2$ is possible on a variety of substrates (e.g. GdScO$_3$, TiN, Si) resulting in different growth relationships.
In particular, being dependent on the nature of the substrate surface and on the growth procedure,  the direction of growth can  sometimes be far from trivial.
In the depicted framework, rather than focusing on a specific growth relationship, we therefore take a more general approach and analyze the effects of a set of elementary strains applied to the primitive cell.
Specifically, we  analyze the case of volume-conserving normal strain directed along the orthorhombic primitive vectors for six strain values $\epsilon$: -3\%,-2\%,-1\%,0\%,1\%,3\%. 
We note that the combination of these simpler cases  allows to deduce how strain influences the hafnia stability, even when directed along non-trivial directions.
Due to the $Pca2_1$ orthorhombic symmetry, the strain affects the distortion modes in different ways depending to the direction of application.
As such, each $\epsilon$ has to be applied along each primitive vector.

\begin{figure}[b]
\centerline{
\includegraphics[angle=-90,width=0.52\textwidth]{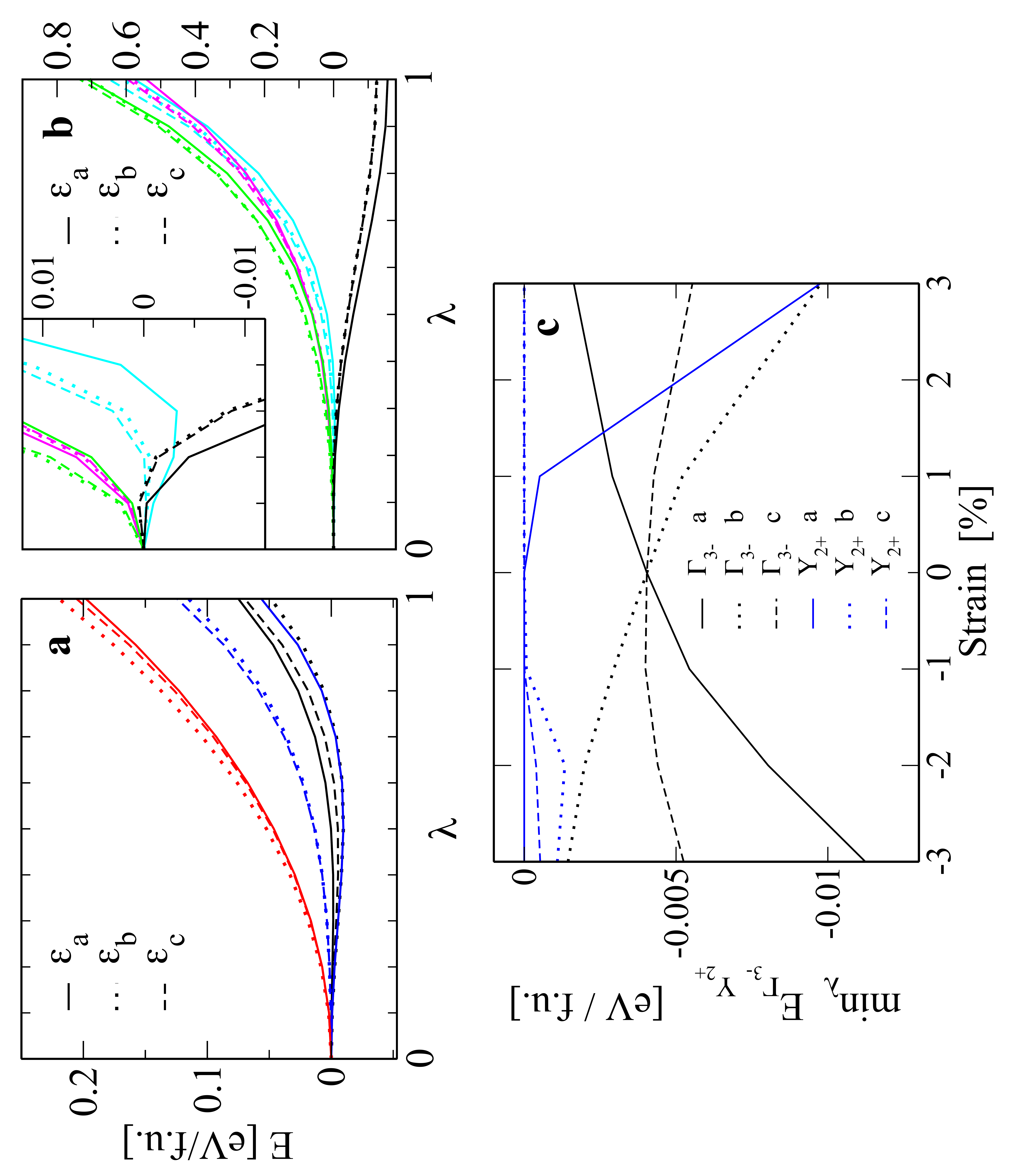}
}
\caption{ \label{strain_plus_min} Panel a,b: {\it ab-initio} energies (in eV/f.u.) of the single symmetry-distortion modes (panel a), along with their couplings (panel b), as a function of the generalized distortion coordinate $\lambda$, when a 3\% tensile strain is applied along \textbf{a}. 
The inset in panel b reports a zoom of the energies for $\lambda$ in $[0.0,0.5]$.  
The colors choice follows the one int Fig.\ref{panel}: $\Gamma_{3-}$ in black, Y$_{2+}$ in blue, Y$_{4-}$ in red (panel a); $\Gamma_{3-} \cup \rm Y_{2+} $ in light-blue, $\Gamma_{3-} \cup \rm Y_{2+} $, $\Gamma_{3-} \cup \rm Y_{4-} $ in purple and $Y_{2+} \cup \rm Y_{4-} $ in green (panel b). The excitation of the three modes is shown in black in panel b. Each path consists of 10 intermediate structures. 
Panel c: the dependence of energy minima of the modes $\Gamma_{3-}$, in black, and Y$_{2+}$, in blue, on the applied strain. In each panel, the application direction is represented by a different line kind.
}
\end{figure}
\subsection{Uniaxial strain effects}
Let us first consider the evolution of the single modes. We refer to panel a of Fig.\ref{strain_plus_min} for an example of the effects of a 3\% tensile strain acting along the different Cartesian directions on the single modes. 
Each color corresponds to a different mode and each line kind refers to a different application direction.
The {\it ab-initio} total energies computed along the distortions reveal how the polar mode $\Gamma_{3-}$ and the pseudo-rotation $Y_{4-}$ maintain their character ({\em i.e.} soft and hard, respectively) independently on the applied $\epsilon$. The only response is a variation of the mode softness or hardness with a different strain modulus and direction of application.
The effects on the second pseudo-rotation, instead, are more complex, as can be seen from the panel c of Fig.\ref{strain_plus_min}, where
the dependence of the Y$_{2+}$ energy minimum on $\epsilon$ is reported. The trend of the $\Gamma_{3-}$ instability is also reported as a comparison. 
The Y$_{2+}$ mode becomes unstable when tensile strain is applied along \textbf{a} and when compressive strain is applied along \textbf{b} and \textbf{c}, whereas the other strains do not induce a softening of the mode. 
The comparison between panels a of Fig. \ref{panel} and Fig. \ref{strain_plus_min} gives an example of the evolution of Y$_{2+}$ for the specific case of 3\% tensile strain applied along \textbf{a}.
A negative $\epsilon$ directed along \textbf{b} and \textbf{c} induces a very weak instability in the pseudo-rotation, as shown by the shallow minimum of the {\it ab-initio} energy profile ({\em i.e.} below $5$ meV/f.u.).
On the other hand, the instability is strongly enhanced by positive \textbf{a}-directed strain: the energy gain $\approx 0.039$ eV/f.u. is comparable with the one of the polar-mode when tensile strain is applied along \textbf{b}.
Thus, essentially Y$_{2+}$ becomes an extra soft mode under certain conditions.
The evolution of this pseudo-rotation under external mechanical actions is strongly asymmetric with respect to the sign of the strain.
Such an uneven response is not specific of the Y$_{2+}$ mode, but it is common to the polar mode, with some differences though. 
Indeed, asymmetric trends in $\Gamma_{3-}$ energy profile are only detected along \textbf{a,b}-directed strain; on the other hand, a \textbf{c}-directed strain tends to lower the energy minimum, in both compressive or tensile cases, resulting in a symmetric profile. 
Moreover, $\Gamma_{3-}$ remains unstable under all the considered strain conditions, as confirmed by the finite energy minima. 
The strong dependence of Y$_{2+}$ on strain shows how an external mechanical action could give rise to non trivial transformation of the single modes.
Nonetheless, the action of strain does not change the need for a combination of the three single modes to complete the transition to the ferroelectric phase.
Panel b of Fig.\ref{strain_plus_min} reports the effects of tensile 3\% strain on the distortions coupling.
Specifically, none of the combinations of two single modes is capable of driving the parent structure towards the $Pca2_1$.
Indeed, the hybrid polar mode stemming from the coupling of the pseudo-rotations retains its stable character for all the considered strains, independently on the application direction.
This is also the case for the  $\rm Y_{4-}\,\cup\,\Gamma_{3-}$ coupling, where the softness of the strained polar mode does not balance the hardness of the Y$_{4-}$ rotation.
Instead, strain destabilizes the combination $\rm Y_{2+}\,\cup\,\Gamma_{3-}$, as can be seen in the inset of panel b in Fig.\ref{strain_plus_min}. 
The largest instability arises for \textbf{a}-directed tensile strain at its highest value considered: $\lesssim 3.5$ meV/f.u. . 
The other strain combinations either give rise to smaller minima or do not change the character of the combined distortion at all.
Despite the insurgence of these instabilities, the coupling $Y_{2+}\,\cup\,\Gamma_{3-}$ can not guide the transition, as shown in panel b of Fig.\ref{strain_plus_min}.
Instead, as it was the case for the unstrained system, the simultaneous activation of the three distortions is required to reach the ferroelectric phase. 
\begin{table}[]
    \centering
    \caption{Difference in total    energy $\Delta$ (in eV per formula unit) between the $Ccce$ and the $Pca2_1$ structures, when a strain $\epsilon$  is applied along \textbf{a},\textbf{b},\textbf{c}.}
    \label{Delta_tab}
    \vspace{0.5cm}
    \begin{tabular}{c c c c}
    \toprule
    \toprule
        $\epsilon [\%]$ & $\Delta_{\bf{a}}$ [eV/f.u.] & $\Delta_{\bf{b}}$ [eV/f.u.] & $\Delta_{\bf{c}}$ [eV/f.u.]\\
    \midrule
    -3  &  -0.113 & -0.152 & -0.143 \\
    -2  &  -0.115 & -0.141 & -0.136 \\
    -1  &  -0.120 & -0.132 & -0.130 \\
     0  &  -0.126 & -0.126 & -0.126 \\
     1  &  -0.135 & -0.123 & -0.124 \\
     3  &  -0.156 & -0.125 & -0.125 \\
    \bottomrule
    \bottomrule
    \end{tabular}
\end{table}
\begin{figure}[t]
\centerline{
\includegraphics[angle=0,width=8cm]{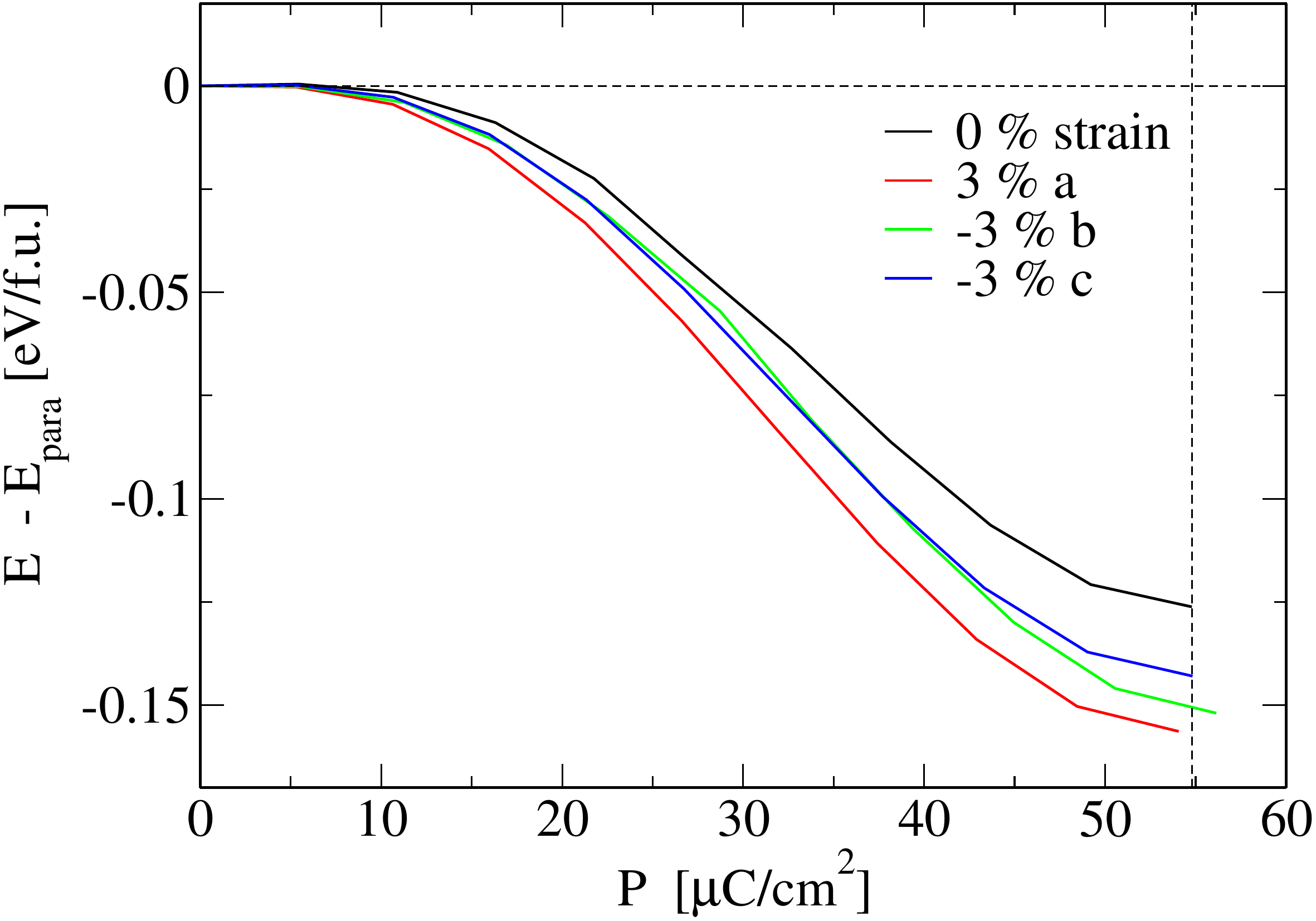}
}
\caption{ \label{pol_toten} 
  Difference in {\it ab-initio} energies between the $Ccce$ and the $Pca2_1$ as a function of the total polarization along the distortion. Only the trends for the most favourable strain states reported in table \ref{Delta_tab} are shown. 
}
\end{figure}

Table \ref{Delta_tab} reports the total-energy difference $\Delta$ between the $Ccce$ and the $Pca2_1$ for the different strain states.
The polar phase becomes progressively more stable as tensile strain increases along the \textbf{a} direction. Instead, it gains stability for increasing compressive strain when applied along \textbf{b} and \textbf{c}.
Since the applied strain is volume-conserving, compressive strain along \textbf{b} and \textbf{c} correspond to an expansion along \textbf{a}.
Thus, we infer that, in order to further stabilize the ferroelectric transition, an experimental strain state should involve the stretching of the \textbf{a} primitive vector from its minimum-enthalpy value.
However, the  stabilization connected to strain effects may come, in some cases, with a loss in the electric polarization.
Indeed, Fig.\ref{pol_toten} reports the relation between the difference in total energy between the paraelectric and the polar phase and the electric polarization arising along the distortion from the $Ccce$ to the $Pca2_1$ phase.
We report only the most favourable strain states from table \ref{Delta_tab}: each strain condition further stabilizes the transition with respect to the non-strained case (black line).
The comparison of the ferroelectric phase polarization under different strain conditions, shows how the enhanced stability corresponds to a slight variation of the electric polarization. 
This is the case of tensile strain applied along \textbf{a} where the polarization becomes $54.07\,\mu\rm C/cm^2$,
to be compared with its value $54.78\,\mu\rm C/cm^2$ under zero strain applied (vertical dashed black line). 
Instead, compressive strain along \textbf{b} and \textbf{c} corresponds to a slight increase to $56.14\,\mu\rm C/cm^2$ and $54.83\,\mu\rm C/cm^2$, respectively.
\begin{figure}[t]
\centerline{
\includegraphics[angle=0,width=8cm]{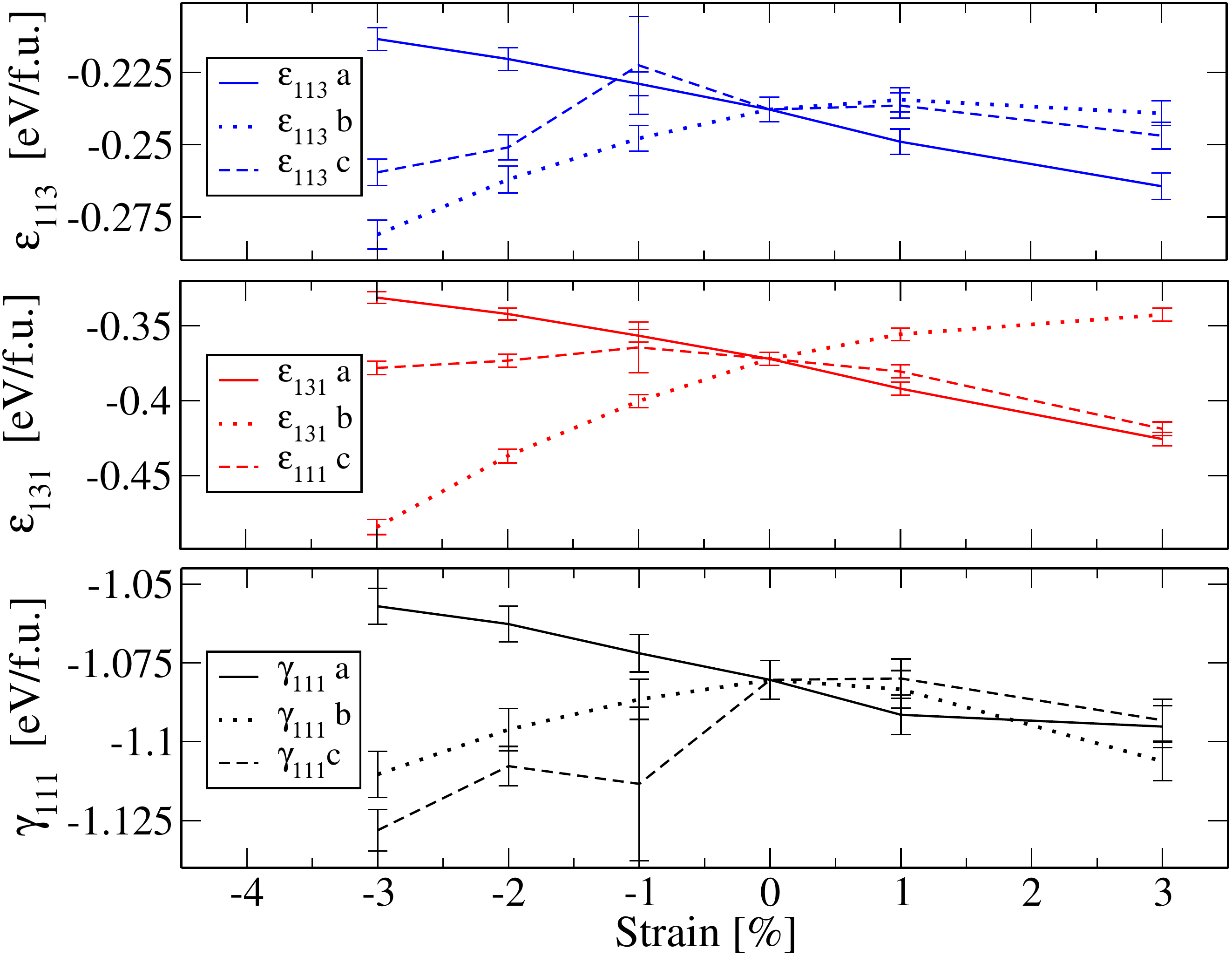}
}
\caption{ \label{param_trend} 
  Strain dependence of the three leading coefficients in eq.\ref{landau}. Top, middle and lower panels show the $\epsilon_{113},\, \epsilon_{131},\, \gamma_{111}$ coefficients, respectively. The application direction of the strain is represented by a different line kind (see legend).
}
\end{figure}

In order to further investigate how strain affects different terms in the Landau free energy, we extract the parameters describing the free energy landscape surrounding the high-symmetry $Ccce$ under different strain states.
The comparison of the extracted coefficients shows that the relative weight of the different terms in defining the Landau-energy profile does not change as the strain varies, at least for the four largest  (in absolute value) coefficients.
Instead, the relative intensity of the remaining coefficients depends on the applied strain.
%
%
Fig.\ref{param_trend} reports the trend of $\gamma_{111},\, \epsilon_{131},\, \epsilon_{113}$ with the errors of the fit as a function of the applied strain. 
In each panel, the continuous, dotted and dahsed lines represent \textbf{a}-directed strain, \textbf{b}-directed, and \textbf{c}-directed, respectively; the Landau coefficients are given in eV/f.u..
The top panel represents the trend of $\epsilon_{113}$. The parameter shows a clear drop -- increasing its absolute value -- upon increasing the strain along \textbf{a}, whereas the opposite trend is observed increasing strain values up to the zero-strain condition along \textbf{c} and \textbf{b}. A further increase of strain along these directions corresponds to a rather flat behaviour of the parameter.
The middle panel shows the trend of $\epsilon_{131}$. When increasing the strain  along \textbf{a} and \textbf{c}  the parameter changes towards larger negative values, whereas increasing the strain along \textbf{b} from compressive to tensile corresponds to a significant energy increase $\gtrsim 0.1$ eV/f.u.. 
Even with the significant changes experienced by $\epsilon_{131}$ and $\epsilon_{113}$ under certain strain conditions, $\gamma_{111}$ remains the largest (in absolute value) parameter for any considered strain by at least 0.5 eV/f.u. . 
%
Thus $\gamma_{111}$ remains the parameter leading the transition in any analyzed configuration. 
As reported in the bottom panel of Fig.\ref{param_trend}, negative and positive strain along \textbf{a} respectively increase and decrease the trilinear coefficient with respect to its value at zero strain. 
Instead, \textbf{b} and \textbf{c}-directed strains produce almost symmetric effects for compressive and tensile, corresponding in both cases to an energy gain with respect to the zero strain condition. 
The fourth three-modes coupling parameter $\epsilon_{311}$ is almost equivalent in values and trend to $\epsilon_{113}$ and it is therefore not shown. 
%
%
It is worth noticing that, in agreement with the change in character of the $\rm Y_{2+}$ rotation, the only parameter changing sign when moving from compressive to tensile strain is $\beta_{\rm Y_{2+}}$, whereas the other second-order coefficients retain their sign. 
Also, the corresponding fourth-order parameters keep their sign fixed: specifically $\delta_{\rm Y_{2+}}$ remains positive.

\subsection{Experimental strain}

We can now combine the effects of 
the considered elementary strains to reproduce an experimentally relevant configuration. 
The strain state of the film originates from the local lattice mismatch and relative orientation of the film with respect to the substrate.
In the case of hafnia, the relative orientation is complicated by non-trivial growth directions. Accordingly, the strain applied to the primitive cell of hafnia will not be directed along its primitive lattice vectors.
We focus here on HfO$_2$ grown on LaSrMnO$_3$ substrate, which has been reported to allow for the deposition of high-quality mono-crystalline films \cite{Noheda_2019,Estandia_2019}. 
Specifically we focus on the work of Estandìa et al. \cite{Estandia_2020}, 
where domain matching epitaxy (DME) is recognized as the growth mechanism of HfO$_2$ [111] on LaSrMnO [001].
%
%
DME allows to accomodate high strains given by the large misfit between the substrate and the film, through the formation of local domains separated by dislocations planes detectable by X-Ray Diffraction (XRD).
This mechanism allows for the minimization of the effective strain:
\begin{equation}
\label{eq_eff_strain}
\rm f^*=(n\cdot d_{LSMO}-m\cdot d_{HfO_{2}})/n\cdot d{HfO_{2}}
\end{equation}
where $\rm d_{LSMO}$ and $\rm d_{HfO_2}$ represent respectively the distance of the n-LSMO layers and m-HfO$_2$ layers forming the (m,n) domain, measured in the interface plane.
Different domains appear in the hetero-structure in order to accomodate the residual lattices mismatch.
Through XRD and STEM measurements, Estandìa et al. recognize four possible orientation of the hafnia supercell equivalent to the one defined by the relations: [-211]$_{\rm HfO_{2}}$ $\parallel $ [110]$_{\rm LSMO}$ and [0-22]$_{\rm HfO_{2}}$  $\parallel$ [-110]$_{\rm LSMO}$.
The domains appearing most frequently along this orientation are: the (9,10) and the (3,2) respectively along the [-211]$_{\rm HfO_{2}}$ and the [0-22]$_{\rm HfO_{2}}$.
Hereafter we focus on this specific combination.
We decompose the strain state experienced by this domain on volume-conserving normal strains applied to the HfO$_2$ primitive cell, named $\boldsymbol{{\rm \varepsilon_a}},\boldsymbol{{\rm \varepsilon_b}},\boldsymbol{{\rm \varepsilon_c}}$.
To this end, we convert the tensors representing these strains to the experimental basis set identified by the Miller indices $\{[-2,1,1],[0,-2,2],[1,1,1]\}$, and we look for a combination of the elementary strains $\boldsymbol{{\rm \varepsilon'_a}},\boldsymbol{{\rm \varepsilon'_b}},\boldsymbol{{\rm \varepsilon'_c}}$ to reproduce the stress state identified by Estandìa et al.
(see Appendix \ref{Methods2} for the detailed derivation)
%
According to Estandìa the HfO$_2$ domain (9,10)-(3,2) strain state is defined by the follwing conditions:
\begin{equation}
\label{exp_strain_conditions}
    \varepsilon'_x = 0.03\% \quad \varepsilon'_y = 5.3\% \quad \sigma'_{xy} = 0\% \, .
\end{equation}
where $\hat{x}=[-2,1,1]$ , $\hat{y}=[0,-2,2]$ and $\hat{z}=[1,1,1]$.
Combining the three volume-conserving normal strains written in the experimental reference  frame   (eq. \ref{strain_exp_volcons})  as follows $\boldsymbol{{\rm \varepsilon'}} = \boldsymbol{{\rm \varepsilon'_a}}-\boldsymbol{{\rm \varepsilon'_b}} -\boldsymbol{{\rm \varepsilon'_c}}$, we obtain a tensor $\varepsilon'$ that respects the conditions expressed in equation \ref{exp_strain_conditions}.
Indeed, the third condition reduces to:
 $\epsilon_b=\epsilon_c$;
by approximating the first condition to $\epsilon'_x\approx0$, we obtain: $\epsilon_a=-\epsilon_b \,.$
The ``experimental" strain tensor therefore becomes:
\begin{equation}
\boldsymbol{{\rm \varepsilon'}} = 
\begin{pmatrix}
    0 & 0 & -\frac{2}{\sqrt{2}}\epsilon_b \\
    0 & -\frac{3}{2}\epsilon_b & 0 \\
    -\frac{2}{\sqrt{2}}\epsilon_b & 0 & 0 
\end{pmatrix}.
\end{equation}
Thus we found a combination of elementary strains applied to the primitive cell which corresponds to the experimental strain state.
Furthermore, in the context of a symmetry distortion mode analysis, we can show how this combination corresponds to a decrease of the Landau free energy with respect to the zero-strain condition. 
From the first and the second conditions in equation \ref{exp_strain_conditions} we know that $\varepsilon'_x\approx0$ and $\varepsilon'_y>0$. 
Thus, $\varepsilon_a>0$ and $\varepsilon_b,\varepsilon_c<0$.
Since $\varepsilon'_y=5.3\%$ we find: $\varepsilon_b=\varepsilon_c=-3.53\%$ and  $\varepsilon_a=3.53\%$.
By comparing these results with Fig.\ref{param_trend}, we recognize how the detected elementary strain combination lowers the leading coefficients of the free energy with respect to the unstrained case.
This condition corresponds to a further stabilization of the ferroelectric $Pca2_1$ phase with respect to the paraelectric phase.
%
%
%
We stress that this decomposition only applies to the specific strain state reported for the domain detected most frequently in Ref. \cite{Estandia_2020}. 
As such, our simplified analysis has the only intent of showing how it is possible to analyze the role of the substrate on the enhanced stability of HfO$_2$ ferroelectric phase.

\section{Conclusions}
In this paper we addressed the nature of the ferroelectric transition of HfO$_2$ through an analysis of the symmetry-allowed distortion connecting the tetragonal high-symmetry phase to the polar orthorhombic phase. %
We emphasize the need for the simultaneous activation of three distortion modes in order to achieve the polar phase. 
The first distortion $\Gamma_{3-}$ is a polar unstable mode at the zone center, whereas the other two Y$_{2+}$ and Y$_{4-}$ are hard modes having a non-polar and anti-polar character, respectively.
The two hard modes become polar only when considered as coupled, giving rise to a hybrid mode that, however, retains its hard character.
The term dominating the Landau free energy expansion is the trilinear coupling, with the fifth-order terms acting as higher order corrections. 
The lack of a clear leading order parameter raises the question on whether the energetics is affected by different factors, such as different growth conditions.
For this reason we analyzed how the strain affects the depicted framework.
Our results show that the need for the coupling of the three modes remains unaltered, independently on the specific external mechanical action. The complex dependence on strain appears in the change of nature of the Y$_{2+}$ mode, which becomes unstable under certain conditions. Nonetheless, as for $\Gamma_{3-}$, this instability can not induce the complete transition by itself.
The fit of the free energy strengthens the idea that the simultaneous excitation of the three modes, along with their coupling, are necessary for the full transition to occur.
Finally, we applied this framework to the realistic case of HfO$_2$ grown on LaSrMnO$_3$. This configuration is complicated by DME, proposed in the context of an epitaxial film subjected to a uniform strain.
We considered the domain experimentally detected most frequently by Estandìa et al., but even in this case, our analysis on the effect of the strain f$^*$ is valid only locally.
Nonetheless, though applied with a heuristic approach, the strain decomposition confirms the existence of strain states that favour the transition towards a ferroelectric phase, along with some others that penalize it.
The aim of our analysis here is not to give an exhaustive explanation for the mechanism underlying the ferroelectric transition in HfO$_2$, but is rather meant as a first step towards a general microscopic description of the effect.
Further steps in this direction will be the analysis of the single-modes features to detect the microscopic origin of the energy gain of their trilinear coupling, along with the relative stability of the monoclinic vs orthorhombic phases in strained film-bulk performed through symmetry mode analysis, issues which we leave for future works.

\appendix
\section{Computational details}
\label{Methods}
Density functional theory simulations were performed using the Vienna Ab-initio Simulation Package (VASP) \cite{Kresse_1999}. 
We relaxed the orthorhombic primitive cell within the revised Perdew, Burke, and Ernzerhof functional for solids \cite{Perdew_2008,Perdew_2009} 
until total energy change between to successive self-consistent steps is smaller than $10^{-7}$ eV and forces are smaller than $10^{-3}$ eV$\rm/\AA$. 
The cutoff for the expansion onto the plane waves basis is set to 400 eV.
As for the distortion modes analysis, we compute the energy of structures along the distortion paths, with PBEsol functional and a k-point density of $\approx 4.5\cdot10^3$ points$\rm \cdot\AA^3$. 
The strain is included applying the volume-conservative deformations reported in equation \ref{eq_volume_strain} to the orthorhombic primitive cell. 
We stress that these transformations consists in pure normal strains: the orthorhombic symmetry is thus preserved.
Once the primitive cell is deformed, the internal coordinates are relaxed until the forces are smaller than $10^{-3}$ eV$\rm/\AA$.
The distortion paths employed to perform the fits under different strain conditions are 
built from these relaxed configurations.
The symmetry-distortion analysis summarized in Fig.\ref{pca21_symmetry_tree} is repeated for each different considered strain.
The electric polarization is computed through the Berry-phase method \cite{Vanderbilt_1993,Resta_1994,Spaldin_2012} 
as implemented in VASP.

\section{Strain}\label{Methods2}
The change from the basis identified by the Miller indices $\{[1,0,0],[0,1,0],[0,0,1]\}$ (where we performed the simulations) to the one identified by  $\{[-2,1,1],[0,-2,2],[1,1,1]\}$ is obtained through the transformation: $$ \boldsymbol{{\rm \varepsilon'}} = \boldsymbol{{\rm Q^{T} \varepsilon Q }}$$ where $\boldsymbol{{\rm Q}}$ is the direction-cosines matrix and $\boldsymbol{{\rm Q^T}}$ is its transpose.
Thus $\boldsymbol{{\rm Q_{i,j}}{\rm =  \hat{e}'_i \cdot \hat{e}_j}}$ where $\rm \hat{e}'_i$ is a vector of the supercell basis, and $\rm \hat{e}_j$ a vector of the primitive cell basis both normalized to unity.
%
%
The $\boldsymbol{{\rm Q}}$ matrix is:
\begin{equation}
\label{Qmatrix}
\boldsymbol{{\rm Q}} = 
\begin{pmatrix}
    -\nicefrac{2}{\sqrt{6}} &  \nicefrac{1}{\sqrt{6}} & \nicefrac{1}{\sqrt{6}} \\
    0 & -\nicefrac{1}{\sqrt{2}} & \nicefrac{1}{\sqrt{2}} \\
    \nicefrac{1}{\sqrt{3}} & \nicefrac{1}{\sqrt{3}} & \nicefrac{1}{\sqrt{3}} 
\end{pmatrix}
\end{equation}
Since $\boldsymbol{{\rm \varepsilon}}$ is diagonal, the strain matrix in the ``experimental" reference frame will be:
\begin{widetext}
\begin{equation}
\boldsymbol{{\rm \varepsilon'}} = 
\label{exp_strain_matrix}
\begin{pmatrix}
    \frac{1}{6}(4\epsilon_a+\epsilon_b+\epsilon_c) &  \frac{1}{\sqrt{12}}(-\epsilon_b+\epsilon_c) & \frac{1}{\sqrt{18}}(-2\epsilon_a+\epsilon_b+\epsilon_c) \\
    \frac{1}{\sqrt{12}}(-\epsilon_b+\epsilon_c) & \frac{1}{2}(\epsilon_b+\epsilon_c) & \frac{1}{\sqrt{6}}(-\epsilon_b+\epsilon_c) \\
     \frac{1}{\sqrt{18}}(-2\epsilon_a+\epsilon_b+\epsilon_c) & \frac{1}{\sqrt{6}}(-\epsilon_b+\epsilon_c) & \frac{1}{3}(\epsilon_a+\epsilon_b+\epsilon_c) 
\end{pmatrix}.
\end{equation}
\end{widetext}
Consider now, for instance, a volume conserving normal strain $\varepsilon_a$ directed along the primitive vector \textbf{a}; the primitive vectors transform as:
\begin{equation}
\begin{split}
    \label{eq_volume_strain}
    a &= a_0(1+\epsilon_a) \\ 
    b &= b_0/\sqrt{1+\epsilon_a} \\ 
    c &= c_0/\sqrt{1+\epsilon_a} .
\end{split}
\end{equation}
In the limit of small $\varepsilon_a$ the strain tensor reads:
\begin{equation*}
\boldsymbol{{\rm \varepsilon_a}} = 
\begin{pmatrix}
    \epsilon_a &  0 & 0 \\
    0 & -\nicefrac{\epsilon_a}{2} & 0 \\
    0 & 0 & -\nicefrac{\epsilon_a}{2} 
\end{pmatrix}
\end{equation*}
"small strains" refers here to values up to $\approx$5\% for which the discrepancy between $1/(\sqrt{1+\epsilon})$ and $1-\epsilon/2$ is  $\approx 0.1\%$.
The tensors for strains along \bf{b} \rm and \bf{c} \rm are analogous.
These three strains when acting independently are written in the experimental frame of reference as:
\begin{widetext}
\begin{equation}
\boldsymbol{{\rm \varepsilon'_a}} = 
\begin{pmatrix}
    \nicefrac{\epsilon_a}{2} &  0 & \nicefrac{\epsilon_a}{\sqrt{2}} \\
    0 & -\epsilon_a & 0 \\
    \nicefrac{\epsilon_a}{\sqrt{2}} & 0 & 0
\end{pmatrix}
\qquad
\boldsymbol{{\rm \varepsilon'_b}} = 
\begin{pmatrix}
    -\nicefrac{\epsilon_b}{4} & -\nicefrac{\sqrt{3}\epsilon_b}{4} & \nicefrac{\epsilon_b}{2\sqrt{2}} \\
    -\nicefrac{\epsilon_b}{2\sqrt{2}} & \nicefrac{\epsilon_b}{4} &  -\nicefrac{3\epsilon_b}{2\sqrt{6}} \\
    \nicefrac{\epsilon_b}{2\sqrt{2}} & -\nicefrac{3\epsilon_b}{2\sqrt{6}} & 0 
\end{pmatrix} 
\qquad
\label{strain_exp_volcons}
\boldsymbol{{\rm \varepsilon'_c}} = 
\begin{pmatrix}
    -\nicefrac{\epsilon_c}{4} & \nicefrac{\sqrt{3}\epsilon_c}{4} & \nicefrac{\epsilon_c}{2\sqrt{2}} \\
    \nicefrac{\epsilon_c}{2\sqrt{2}} & \nicefrac{\epsilon_c}{4} &  \nicefrac{3\epsilon_c}{2\sqrt{6}} \\
    \nicefrac{\epsilon_c}{2\sqrt{2}} & \nicefrac{3\epsilon_c}{2\sqrt{6}} & 0 
\end{pmatrix}.
\end{equation}
The combination $\boldsymbol{{\rm \varepsilon'}} = \boldsymbol{{\rm \varepsilon'_a}}-\boldsymbol{{\rm \varepsilon'_b}} -\boldsymbol{{\rm \varepsilon'_c}}$ results in:
\begin{equation}
\boldsymbol{{\rm \varepsilon'}} = 
\begin{pmatrix}
    -\frac{1}{4}(2\epsilon_a+\epsilon_b+\epsilon_c) & \frac{\sqrt{3}\epsilon}{4}(-\epsilon_b+\epsilon_c) & \frac{1}{2\sqrt{2}}(2\epsilon_a-\epsilon_b-\epsilon_c) \\
    \frac{\sqrt{3}\epsilon}{4}(-\epsilon_b+\epsilon_c) & 
    \frac{1}{4}(4\epsilon_a-\epsilon_b-\epsilon_c) &
    \frac{3}{2\sqrt{6}}(\epsilon_b-\epsilon_c) \\
    \frac{1}{2\sqrt{2}}(2\epsilon_a-\epsilon_b-\epsilon_c) & \frac{3}{2\sqrt{6}}(\epsilon_b-\epsilon_c) & 0 
\end{pmatrix}
\end{equation}
\end{widetext}
which can be adapted to represent the experimentally-detected strain state.
To be noticed: there exists a combination that would lead to a diagonal tensor $\boldsymbol{{\rm \varepsilon'}}$, which is: $\varepsilon_b=\varepsilon_c=\varepsilon_a$. 
But in such case we find $\varepsilon'_x=\varepsilon_a$ and $\varepsilon'_y=\varepsilon_a/2$, which can not reproduce the conditions in eq. \ref{exp_strain_conditions}.

\end{document}